\documentclass{knawproc}
\usepackage{epsfig}

\def\deg{\hbox{$^\circ$}}
\def\arcsec{\hbox{$^{\prime\prime}$}}
\def\farcs{\hbox{$.\!\!^{\prime\prime}$}}

\begin{document}

\begin{opening}

\title{Hosts and Alignment Effect in $0.5 < z < 1$ radio sources}

\author{Susan E. Ridgway}
\addresses{%
Astrophysics, NAPL\\ University of Oxford \\ Keble Rd\\Oxford OX1 3RH UK\\
}

\end{opening}


\begin{abstract}

$R$- and $I$- band images with good seeing at the NOT of the hosts of a 
sample of 7C radio galaxies at $z \sim$ 0.7 (with mean 
radio flux 20 times fainter than the 3C sample)
show some evidence of radio-optical alignment.  A quantitative analysis
of the alignment in this sample and in a corresponding 3C sample 
from HST archival data indicates that the percentage of aligned flux may be
lower and of smaller spatial scale
in the faint radio sample.
Studies of the aligned material in $z \sim $1
3C source hosts indicate that the quasar extensions also show some aligned 
flux, though much of it is due to optical synchrotron radiation. 

\end{abstract}


\section{Introduction}

Studies of samples of
high radio luminosity radio galaxies at $z \geq  0.8$ 
show that the position angles of the optical and emission-line 
morphologies tend to 
align with the radio axis (McCarthy et al.\ 1987, Chambers et al.\ 1987).
High-resolution optical imaging with the Hubble Space Telescope
revealed that the aligned material is morphologically 
distributed in a wide variety of
ways (e.g. Best et al.\ 1996, 1997), and is unlikely to be the result 
of a single mechanism. 
Apart from 3C368, in most objects nebular continuum
fails to explain more than a fraction of the aligned light. Scattering
of a hidden quasar nucleus is expected to add to the aligned light,
and the detection of broad, polarised Mg{\sc ii} emission in a number of
these objects (e.g.\ 3C265 and 3C324;
Dey \& Spinrad 1996; Cimatti et al.\ 1996) proves that this must be
an important contributor.  The closely aligned morphologies
seen in the WFPC2 imaging are inconsistent with pure scattering
models, and star formation induced by the passage of the radio jet
has been proposed by many (e.g. Chambers et al. 1990, Best et al. 1996, 
Dey et al. 1997).

How will the alignment effect depend on the radio power of the source
sample?
Some of these alignment mechanisms might be expected to result in
a nearly linear dependence of the amount of aligned optical emission
on the radio luminosity of the sources.
The radio luminosity and
the strength of narrow optical emission lines in radio galaxies
are known to be well correlated (Rawlings et al.\ 1989).
Recently, a correlation of the optical continuum luminosity of steep-spectrum
radio-loud quasars with radio luminosity has also been established
(Serjeant et al.\ 1998). Both of these imply a close relationship between
the optical/UV luminosity of the AGN and the radio luminosity, presumably
via a correlation with the bulk kinetic power of the radio jets
(Rawlings \& Saunders 1991). 

Two imaging studies in the near-infrared, that of Dunlop \& Peacock (1991) and
of Eales et al.\ (1997), found that the ``alignment effect'' was unmeasurably
small in samples of $z\sim 1$ radio sources approximately ten and four times
fainter than 3C respectively. Both these studies were, however,
carried out under
conditions of poor ($>1$\arcsec) seeing, and in the $K$-band, where the
alignment effect is weak even in 3C (Rigler et al.\ 1991). They may not
have been able to constrain how strongly the alignment effect depends
on radio luminosity.

Here, I will discuss the result of
a project done in collaboration with Mark Lacy
at the University of Oxford to analyze alignments
in a sample of low-luminosity radio sources ($\sim$ 20 times fainter than 3C) 
from the 7C North Ecliptic Cap (NEC) survey of Lacy et al. (1993).
I will also discuss some of the alignments seen in 3C
quasars at $z \sim 1$, based on a WFPC2 imaging project done in collaboration 
with Alan Stockton at the University of Hawaii. 

\section{Alignments in a 7C sample of radio galaxies}

The 7C NEC sample of sources was selected to a flux limit of 0.5 Jy
at 151 MHz, and has been spectroscopically identified to a completeness
level of $\sim$ 90\% (Lacy et al. 1993, Lacy et al. in prep).  
As a by-product of a project to study the clustering properties of 
a complete, extended 
(radio size $\theta_{\rm r}>1$\arcsec) $0.5<z<0.82$
sub-sample of these 7C sources (Wold et al. 1996,
Lacy this volume), fairly deep, high image quality $I$ or $R$-band
images of the radio galaxies themselves were obtained. 
These images were 40 min exposures obtained
at the 2.56 Nordic Optical Telescope (NOT) with seeing conditions ranging from 
0\farcs5 to 0\farcs9.  
Spectroscopy of all of these objects has been obtained with ISIS
on the William Herschel Telescope (WHT) as part of the 
sample identification process, with slit positions generally along the
radio axis. 

A visual inspection of the images (Fig.\ 1) reveals a number of objects 
in which the optical morphology appears aligned with the radio axis. The
morphologies of the aligned material
differ: in two cases, the hosts are apparently ellipticals
with the major axis within 30\deg\ of the radio PA, while others
(e.g. 7C1748+6731 and 7C1826+6510)
have multicomponent morphologies along the 
radio axis similar to that seen in the high luminosity 
3C sources. However, in the ``elliptical'' host 7C1745+6415,
the spectrum and consequently the morphology visible in the I-band image
are dominated by emission lines rather than stellar continuum (Fig. 2).
On the other hand,
in 7C1826+6510, which has a secondary component $d$ lying along the
radio axis 
well-separated spatially from the host galaxy $b$, the spectrum of 
$d$ confirms that 
it is probably associated. Its spectrum is dominated by stellar continuum and
shows a 4000\AA\ break and stellar absorption
features with a $\sim$8000 km/sec velocity relative
to the central component $b$. Furthermore, visible in the image is diffuse
material lying along the radio axis. 

Therefore, this moderate redshift, low luminosity 
sample of objects seems to have a number of cases
of the alignment
effect similar to that shown in the $z \sim 1 $ 3C sources. To test whether 
this is a statistically significant result, we attempt to make a quantitative
measurement of the ``alignment effect''
and compare these measurements to those derived for a 
sample of 3C galaxies in the same redshift range with imaging data
available in the HST archive. 

\begin{figure}[h!t]
\includegraphics[scale=0.65,angle=0]
{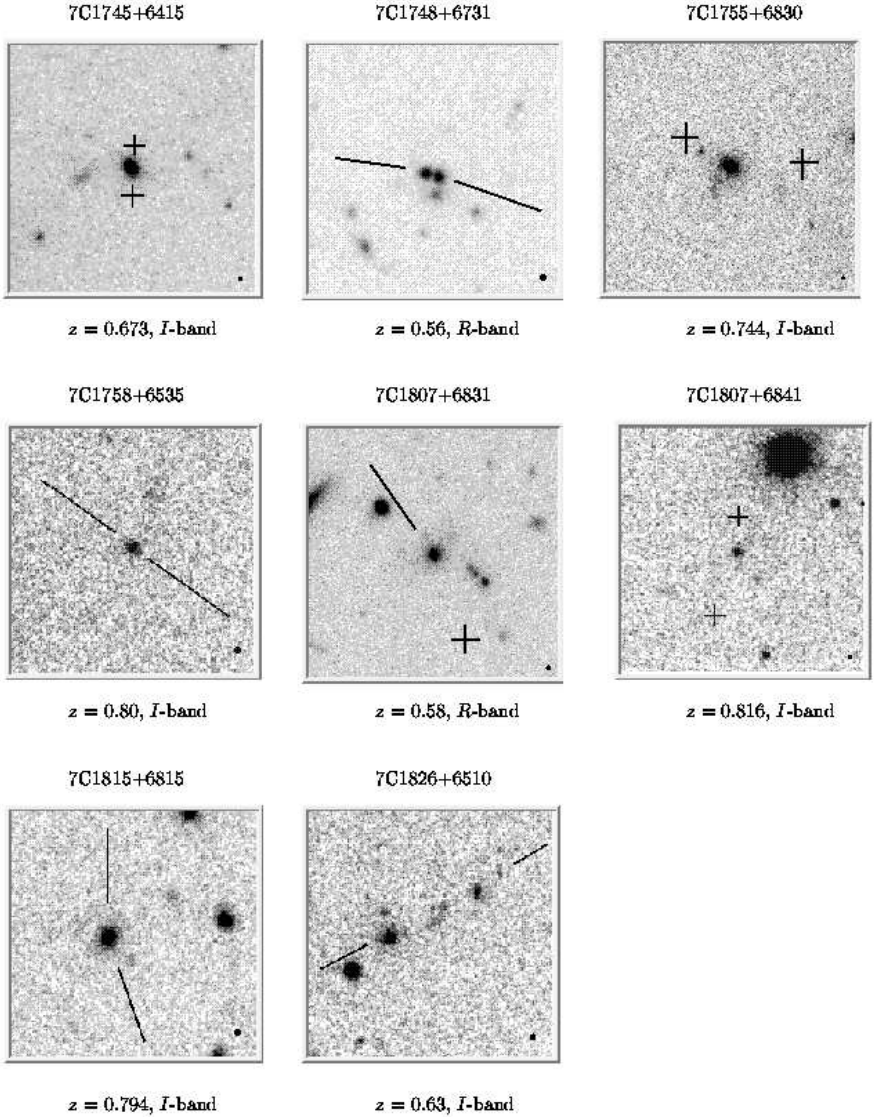}
\caption{The 7C galaxies analyzed in this paper, with N up E left. The images
are 28\arcsec\ square. The black crosses designate the
positions of radio hotspots, while the black lines show directions to the
hot spots if they fall outside the field. The IDs are generally obvious,
except in 7C 1826+6510 where it is the object $b$ to the SE, marked with
the radio lobe line. The aligned object $d$ is marked by the other
radio lobe directional line.}
\end{figure}
\begin{figure}[h!t]
\includegraphics[scale=1,angle=0]
{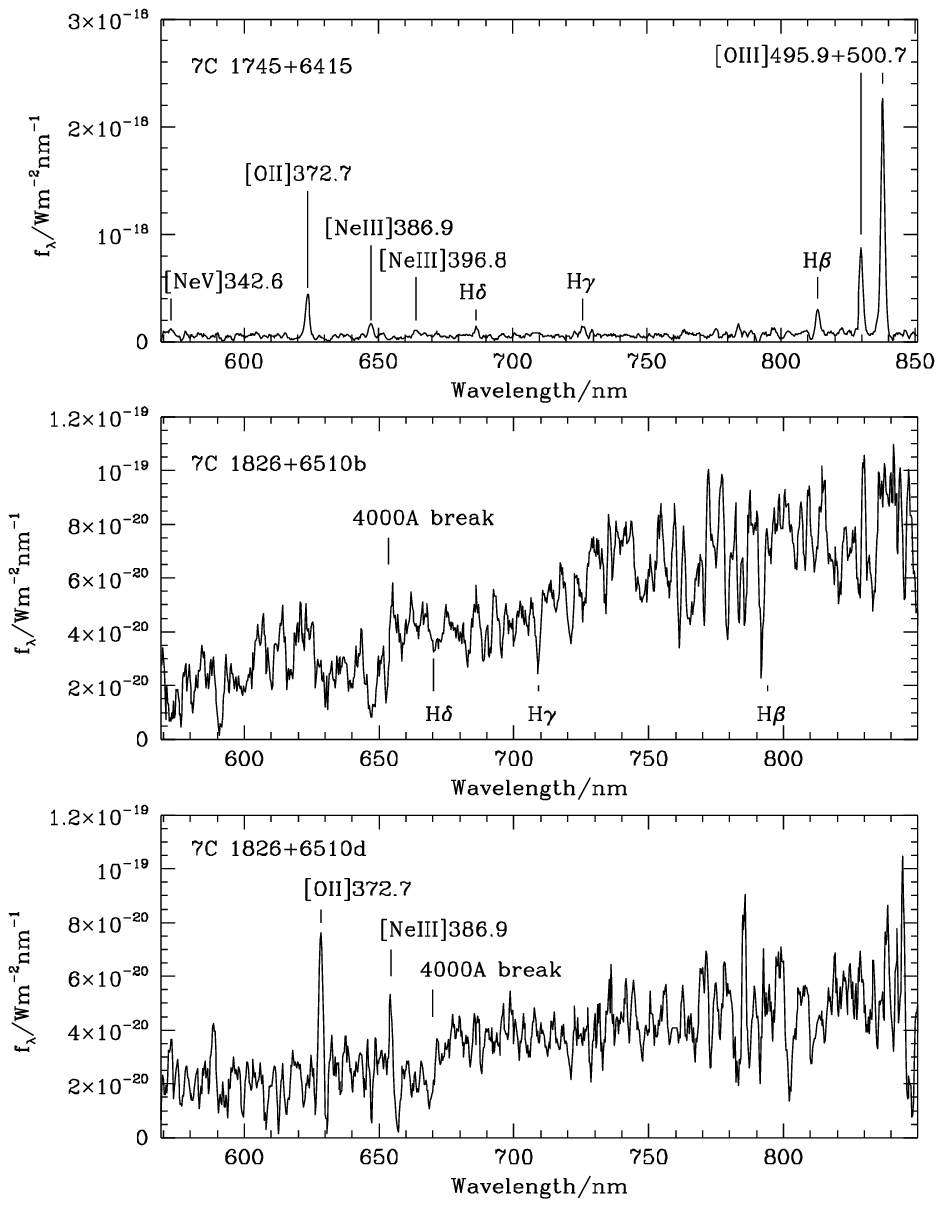}
\caption{Top: WHT spectrum of 7C1745+6415. Middle: WHT spectrum
of 7C 1826+6510 $b$, the radio source identification.
Lower: WHT spectrum  of 7C 1826+6510 $d$, an aligned object.}
\end{figure}

\section{Quantitative Analysis of Alignment Effect}

We have used two methods to measure the alignment of the galaxies in
these samples: first, a standard moment analysis to derive the difference
in the
optical position angle from the radio position angle ($\Delta$ PA)
(Rigler et al. 1992, Dunlop \& Peacock 1994, Ridgway \& Stockton 1997), 
and second, 
a simple derivation of the percentage aligned flux, which should be 
less susceptible to skewing by companions and differences in observational
resolution. To derive this \% aligned flux, we have summed the flux within 
$\pm$45\deg\ of the lines joining the nucleus to each of the radio hotspots,
and subtracted the remaining flux. We exclude from the sums the central portion
of each galaxy within the FWHM of the seeing for the
ground-based data, and for the archival HST data, we exclude
\clearpage
a circular region 
with a diameter (0\farcs5) comparable to the FWHM of the best seeing
ground-based images. 
This \% aligned flux value will be positive for images with more flux
within the cone of the radio axes than without,
and will be negative for objects with more counter-aligned than aligned flux.  

To make intercomparisons between objects in this fairly inhomeogenous dataset,
we must account both for differences in the depth of the observations and
the spread in redshift across the samples. Before analysis, 
we have therefore resampled 
the 7C NOT and 3C WFPC2 
images to a common pixel scale, and then corrected the surface brightnesses
for the
differences in cosmological dimming by normalizing
to the median redshift of the samples ($z$=0.7) (by assuming that $F_{\lambda}
\propto (1 + z)^{-5}$, i.e. that the galaxies' intrinsic spectra are 
approximately flat in $F_{\lambda}$ at these wavelengths). 

A well-known problem with the standard position angle analysis is the
subjectivity of choosing the isophotal cutoff level and the aperture within
which to calculate the moments.
In this paper, we have chosen a single
isophotal cutoff, used for all the images,
which is about 2.5 times the median normalized
sky sigma in these frames for both the position angle analysis and
for the \% aligned flux analysis. 
Inspection of the behavior of the integrated \% aligned flux in increasing 
apertures up to about 10\arcsec\
showed that in general there were two spatial regions which tended to
contain the major flux contributions:
within about 2\arcsec\ ($\sim$15 kpc) and within about 6 \arcsec\
($\sim$50 kpc).
This led us to choose 15 kpc and 50 kpc as the radii of
our standard apertures for our analyses, calculated at the redshift of 
each object using H$_{0}$ = 50 km s$^{-1}$ Mpc$^{-1}$, q$_{0}$ = 0.5 .

We therefore calculate for each object in the sample the $\Delta$PA and the
\% aligned flux as described above, within these two apertures after 
making the isophotal cutoff. For many of these objects,
which were observed longer or were at lower than median redshift,
this standard isophotal cutoff may result in losing obviously well-detected
low surface brightness material: for example, large scale extended aligned
material in 3C 277.2 falls below this cutoff. This is necessary, of course,
to make the comparison on an unbiassed basis. 
We have also calculated
the \% aligned flux in the annulus between 15 kpc and 50 kpc,
and we give in Table 1 the results of these analyses for both samples. 
Also given are the unnormalized aperture magnitudes for the objects.

\begin{table}
\caption{Results of alignment analysis}
{\small
\begin{tabular}{lccccccc}
Name &\% Aligned &\% Aligned& \% Aligned & $\Delta$PA & $\Delta$PA & m(AB) & m(AB)\\
     &   Flux   &  Flux    &  Flux   &           &           &       &\\
     & (15 kpc) & (50 kpc) & (15 -- 50 kpc) &  (15 kpc) & (50 kpc)& (15kpc) & (50 kpc)\\
& 	& 	& 	& 	& 	& 	&\\
3C34   & 4.1 $\pm$  1 &$-$44 $\pm$ 1 & $-$68.1 $\pm$ 1 & 20 & 73 & 20.0 & 19.7 \\
3C41   & $-$6.9 $\pm$ 1 &  16 $\pm$ 1 & 96.3 $\pm$ 3 &  88 & 22 & 20.9 & 20.4 \\
3C172  & 10.6 $\pm$ 1 & 37.2 $\pm$ 1 & 44.8 $\pm$ 1 & 12 & 23 & 19.9 & 18.9\\
3C226  & 24.8 $\pm$  1 &   17.1 $\pm$ 1 & $-$8.5 $\pm$ 2 &  2.7 & 36 & 20.9 & 20.3\\
3C228  & $-$9.9 $\pm$  2 &  - & - & 62 & - & 19.8 & - \\
3C247  & $-$7.6 $\pm$ 1 & 41.0 $\pm$ 1 & 89.6 $\pm$ 1 &  74& 14 &20.2 & 19.1  \\
3C265  & 7.8  $\pm$ 1 &   15.4 $\pm$ 1 & 23.5 $\pm$ 1 &  19 &11  & 20.1 & 19.3 \\
3C277.2 & 22.1 $\pm$  1 & 3.2 $\pm$ 1 & $-$90.0 $\pm$ 1 & 11 &  75 & 20.6 & 20.2 \\
3C337  & 25.2 $\pm$  1 & 55.4 $\pm$ 1 & 79.7 $\pm$ 1 & 2.5 & 10 & 20.7 & 20.2\\
3C340  & 15.8 $\pm$  1 &  19.4  $\pm$ 1  & 92.9 $\pm$ 7 & 13 & 18 & 21.1 & 20.8\\
3C441  & 18.6 $\pm$  1 &  27.5$\pm$ 1 & 33.8 $\pm$ 1 &  17 & 7 & 20.8 & 19.9\\
Median 3C&      11  & 18 & 39   & 17   & 20 & 20.6 & 20.1 \\

  &              &              &      & &     &     &    \\
7C1745+6415&  6.5 $\pm$ 1 & 6.5 $\pm$ 1 & 0 $\pm$ 0 & 35 & 35 & 19.7 & 19.5\\
7C1748+6731&  100 $\pm$  2& 100 $\pm$ 2 &  0 $\pm$ 0 & 1.5 & 1.5 & 21.1 & 20.4 \\
7C1755+6830&  12.2 $\pm$   1&  8.8 $\pm$ 1 &$-$6.1 $\pm$ 3 &  22 & 4  & 21.0 & 20.5\\
7C1758+6535&   $-$3.7 $\pm$   7&  $-$13.4 $\pm$ 6 & $-$47 $\pm$ 15&  2 & 72 & 21.4 & 20.9 \\
7C1807+6831&   15.1 $\pm$ 1&   24.4 $\pm$ 1   & 81 $\pm$ 4 & 11 & 35 & 20.6 & 19.9 \\
7C1807+6841&   $-$9.2 $\pm$   3&   $-$11.5 $\pm$ 2 & $-$12.3 $\pm$ 2 & 17 & 70 & 21.6 & -    \\
7C1815+6815&  6.8 $\pm$ 2&   7.3 $\pm$ 2 & 16.9 $\pm$ 10 & 28 & 25 & 20.2 & 19.7 \\
7C1826+6510&   10.3 $\pm$   3&    4.5 $\pm$ 3   &$-$11.9 $\pm$ 6 &  11 & 22 & 20.1 & 19.5 \\
 Median 7C  &9          & 7  & $-$6         &  14 & 30 & 20.8 & 19.9 \\
\end{tabular}
}
\footnotesize
(1) The errors given are the statistical errors based on the contribution of
sky noise to each sum.
(2) The ``0 $\pm$ 0'' values mean that no flux was left above
the isophotal cutoff in those annuli.
(3) The 3C 228 image was a very short single PC exposure; only the inner portion
was used since all CRs were removed by hand.
\end{table}

\section{Results of Analysis}
 
We give in Figures 3 and 4 histograms of the results of the $\Delta$PA and \%
aligned flux analysis within the 15 and 50 kpc apertures for the 7C and 3C
samples. The \% aligned flux median values given in
Table 1 seem to indicate a comparable degree of alignment in the 7C as in
the 3C sample, at least at the 15 kpc radius.
In the region between 15 kpc and 50 kpc,
however, the 3C sample seems more aligned.

To determine whether these results indicate a significant alignment
in the 7C and 3C samples, we have made a number of statistical tests,
the results of which we give in Table 2. 
First we determine whether the $\Delta$PA values are consistent
with random orientation; a Kolmogorov-Smirnov (K.S.) test 
of the values versus a uniform distribution (Table 2, lines 1 \& 2)
shows
that at 15 kpc both the 7C and the 3C sample have a probability of 
less than 1\% of being randomly oriented.
The same test for the 50 kpc aperture is less conclusive for both
samples, as might be expected from an increased slewing of the PA by 
companion objects, but particularly for the 7C sample (line 3).

What about the \% aligned flux values? 
We have made a K.S. test to determine whether the 7C and 3C \% aligned
flux distributions differ significantly, and find that in the 15 kpc aperture,
there is no significant difference (Table 2, line 5).
Within the 50 kpc aperture, and particularly in the 15 -- 50 kpc annulus,
we find that the probabilities of having such different distributions 
by chance is significantly reduced (Table 2, lines 6,7,8). 

From these tests we can conclude that at least at the 15 kpc scale the 7C
sample we have observed seems to exhibit a significant alignment effect. 
At the larger 50 kpc scale, while the 3C sample continues to show a similar
or perhaps increasing amount of aligned flux, the 7C sample is less well
aligned. 
\begin{figure}[!t]
\includegraphics[scale=0.55,angle=0]
{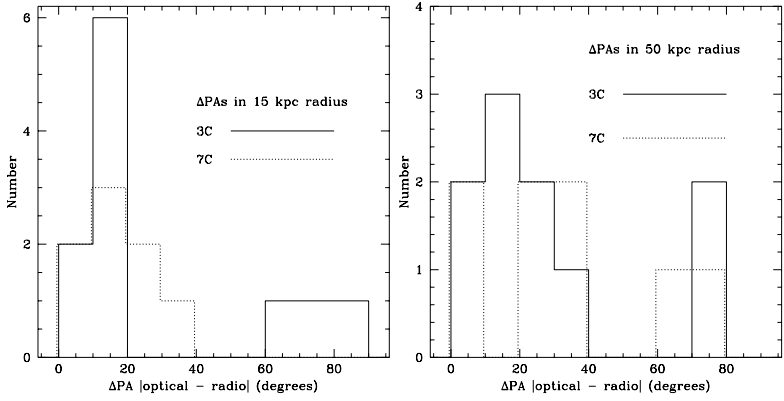}
\caption{Left: $\Delta$PAs (difference between the optical and radio axes)
for the 3C and 7C galaxies within the 15 kpc aperture.
Right: $\Delta$PAs for the two samples within the 50 kpc aperture.}
\end{figure}
\begin{figure}[!t]
\includegraphics[scale=0.55,angle=0]
{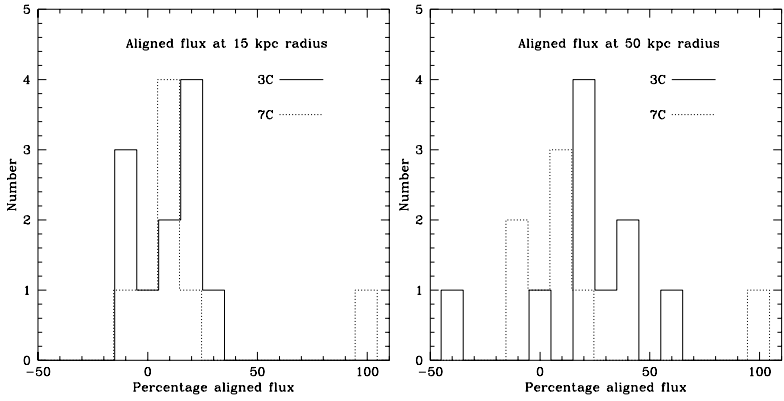}
\caption{Left: Aligned flux percentage
for the 3C and 7C galaxies  within the 15 kpc
aperture.  Right: Aligned flux percentages for the two samples
within the 50 kpc aperture.}
\end{figure}

\begin{table}
\caption{Summary of results of statistical tests}
{\small
\begin{tabular}{lllll}
Number & Distributions tested & Type of Test & Result & Probability\\

1 & 7C $\Delta$PA (15 kpc) vs. uniform & K.S. (One-sided) & D = 0.617 & 0.002 \\
2 & 3C $\Delta$PA (15 kpc) vs. uniform & K.S. (One-sided) & D = 0.505 & 0.004 \\
3 & 7C $\Delta$PA (50 kpc) vs. uniform & K.S. (One-sided) & D = 0.361 & 0.193 \\
4 & 3C $\Delta$PA (50 kpc) vs. uniform & K.S. (One-sided) & D = 0.444 & 0.026 \\
& & & & \\
5 & \% Align. Flux (15 kpc): 7C vs. 3C& K.S. (Two-sided) & D = 0.330 & 0.573 \\
6 & \% Align. Flux (50 kpc): 7C vs. 3C& K.S. (Two-sided) & D = 0.550 & 0.095 \\
& & & & \\
7 & \% Align. Flux (15 -- 50 kpc): 7C vs. 3C& K.S. (Two-sided) & D = 0.575 & 0.070 \\
8 & \% Align. Flux (15 -- 50 kpc): 7C vs. 3C& K.S. (One-sided) & D = 0.575 & 0.035 \\

\end{tabular}
}
\end{table}

\section{Discussion}

Though these results are admittedly based on small samples and are not of
great statistical significance, they are quite suggestive.
We seem to have significant alignment at a 15 kpc
scale in a sample of radio galaxies 
whose luminosity is $\sim$20 times fainter than that of the 3C sample.

Two of the most common alignment mechanisms known to operate in the
3C sample are scattering and nebular continuum emission. Both of these
should scale approximately as the emission line strengths, which are
found to scale approximately with the radio luminosity (to about the 6/7
power; Rawlings \& Saunders 1991). 
We should therefore
expect contributions from scattering and nebular continuum emission
to be about a factor of 15 less in the 7C sample than in the 3C sample;
were the percentage alignments this much diminished they would have been 
undetectable. The probable luminosity 
dependence of other mechanisms are less clear, but may scale 
less severely with radio power.
Jet-induced star formation is not well understood, but could in
theory be more effective with lower jet velocities, and 
the ``selection bias'' effect of Eales (1992) might also operate
effectively in low radio luminosity samples.

Of course, even if we are correct in finding almost comparable
percentage alignments in the 7C sample and the 3C sample at $\sim$ 15 kpc,
this does not necessarily translate into ascribing the same alignment 
mechanism to the two samples. 
An important difference seems to be in the scale of the aligned material.

We have shown that in this sample,
the 3C objects are aligned over a large range in scales.
In the extensively studied $z \sim 1 $ 3C galaxies, of which these are probably
a reasonably typical subset (though at slightly lower $z$), it is known 
that there is a variety of
causes for the alignment effect, some of which would work fairly
well at large scales, such as scattering.
In our 3C sample, some galaxies such as 3C 265
have been studied with ground-based polarimetry and are known to be 
quite polarized, and therefore probably have a large scattering
contribution to the aligned light. 
Though we do not have in our 3C sample any obvious examples in which
the very small scale, ``jet path'' aligned morphology (e.g 3C 324)
dominates, this is probably due to the more dominant elliptical
hosts at longer rest wavelengths, 
and this very small scale alignment mechanism may still 
be contributing.  We can probably assume that in this sample
the same wide range of mechanisms determined in the $z \sim 1$ samples
are contributing, and some of these mechanisms 
are obviously effective at scales of $\sim$50 kpc.

Why, then, are the galaxies in this 7C sample aligned? First, we find that the 
alignment is comparable to that of the 3C sample at scales of 
$\sim$15 kpc, but probably not at the larger scales of $\sim$50 kpc.
We have no information, of course, at the very small (0\farcs1) scale 
at which some 3C objects show alignment. Our result, though reasonably
good statistically, is based on a few objects,
which show a fair bit of variation in type of aligned morphology themselves. 
So it is difficult to tell what the likely mechanisms are, and why they
do not seem to be operating at $\sim$ 50 kpc scales. 

Some possible explanations for the scale difference are that the 
alignment mechanisms that operate at the large scales in the 3C are the 
most luminosity dependent, such as scattering or nebular continuum.
In addition, there may simply be less material at larger scales in the 7C
environments to be aligned or to act as a scattering medium.
(We do find on average, with great dispersion,
twice as much flux above the isophotal cutoff in the 3C as in the
7C in the 15 -- 50 kpc annulus). 

Nevertheless, the alignment that we seem to see 
in the 7C sample indicates that some less-luminosity dependent alignment
mechanism, effective at moderately small scales,
may be contributing to the ``alignment effect'' that we see in radio
sources. That the 3C alignments extend to larger scales could mean
that this less-luminosity dependent alignment mechanism is swamped
by other effects, such as scattering, in the more powerful radio
sources. We will discuss these results in more detail in Lacy et al.
(1998).

\section{Alignment effect in 3CR Quasars}

What about the other type of radio source, the quasars? 
In a WFPC2 project to study a sample of $z \sim 1$ 3CR sources,
Alan Stockton and I obtained $\sim R$ images of 5 steep-spectrum quasars
with extended radio structures (Ridgway \& Stockton 1997).
We were able to resolve circumnuclear
optical emission in all 5 of the quasars, and found that 
in 4 cases the optical structures were
aligned closely with the radio structure.
In one quasar, 3C 196, the type of aligned morphology is similar
to that seen in the 3C galaxies.
In the other three, 
the morphological correspondence between the optical and
radio is so close that it is almost certainly optical synchrotron
emission or similar non-thermal process.
3C 2 has an optical synchrotron northern lobe/hot spot,
while the other two, 3C 212 and 3C 245, both have optical synchrotron jets. 
We have been obtaining high-resolution 
MERLIN mapping of the sources, which has allowed us to determine 
radio-optical spectral indices for many of the components unresolvable 
with previous radio mapping. In Fig. 5, we show the WFPC2 image of
3C 212 with an L band MERLIN map overlaid, and in Fig. 6 we show the derived
SEDs for the various components. We find that the spectral indices and 
probable turnovers are typical of those seen in lower $z$ cases
of optical synchrotron where the synchrotron nature has been proved
with optical polarimetry.
That we saw optical synchrotron in 3/5 quasars and 0/5 galaxies
in this small sample is consistent with beaming and the unification hypothesis
of AGNs.
\begin{figure}[h!t]
\includegraphics[scale=0.6,angle=0]
{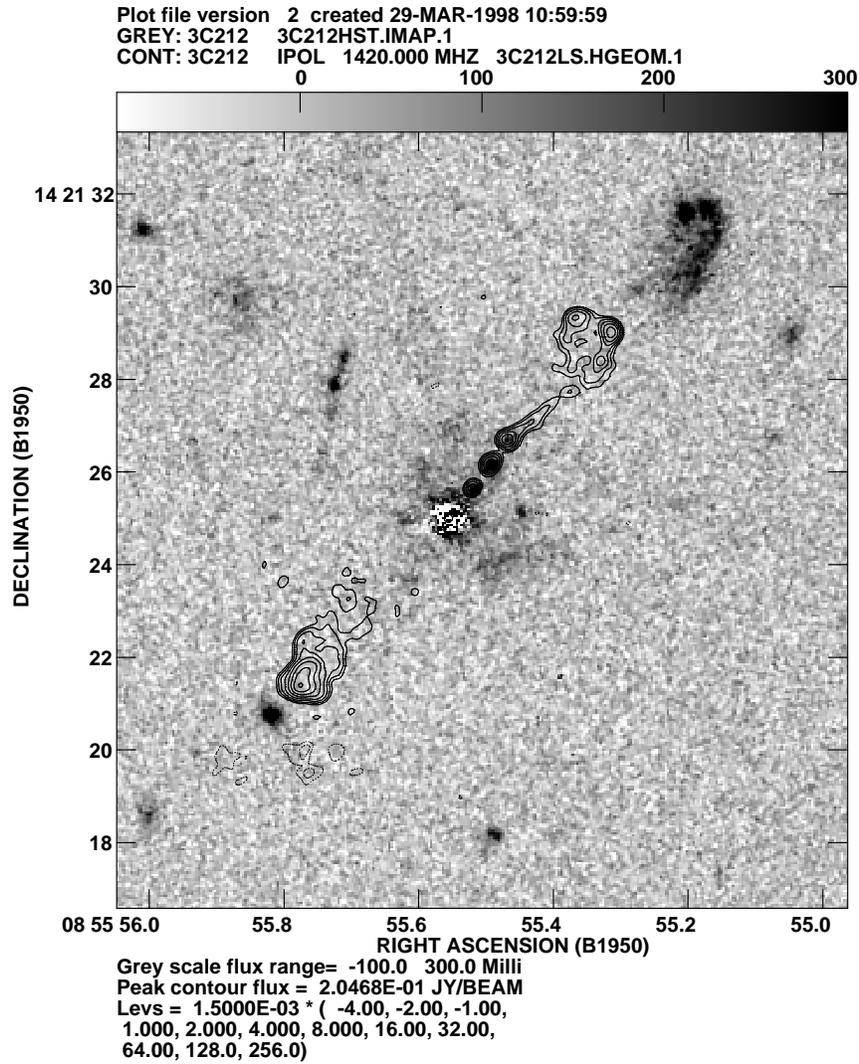}
\caption{WFPC2 image of 3C 212, with the nucleus
PSF-subtracted, and with MERLIN L band map overlaid.
The three optical jet components 
lie to the NW of the nucleus, and correspond exactly in 
position to the radio jet knots seen in the MERLIN image. 
They are referred to as $a$,
$b$ and $c$ in increasing distance from the core. } 
\end{figure}
\begin{figure}[h!t]
\includegraphics[scale=0.4,angle=0]
{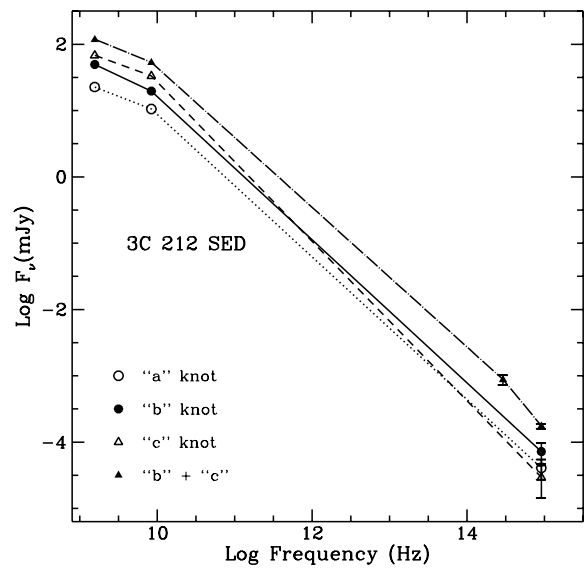}
\caption{The spectral energy distributions of the jet components from
the WFPC2 image, the MERLIN L band map, a VLA X band image, and from
a Keck $K$ band image (which was only able to resolve the sum of the outer
two components, $b + c$.}
\end{figure}

\end{document}